\documentclass{sigchi-ext}
\usepackage[T1]{fontenc}
\usepackage{textcomp}
\usepackage[scaled=.92]{helvet} % for proper fonts
\usepackage{graphicx} % for EPS use the graphics package instead
\usepackage{balance}  % for useful for balancing the last columns
\usepackage{booktabs} % for pretty table rules
\usepackage{ccicons}  % for Creative Commons citation icons
\usepackage{ragged2e} % for tighter hyphenation
\usepackage{amssymb}

\def\plaintitle{Applying the Persona of User's Family Member and the Doctor to the Conversational Agents for Healthcare} 
\def\emptyauthor{}
\def\plainkeywords{Persona; Chatbot; healthcare}

\title{Applying the Persona of User's Family Member and the Doctor to the Conversational Agents for Healthcare}

\numberofauthors{5}
\author{%

\alignauthor{%
    \textbf{Youjin Hwang}\\
    \affaddr{Seoul National University} \\
    \affaddr{Seoul, Republic of Korea} \\
    \email{youjin.h@snu.ac.kr}
}
\alignauthor{%
    \textbf{Donghoon Shin}\\
    \affaddr{Seoul National University} \\
    \affaddr{Seoul, Republic of Korea} \\
    \email{ssshyhy@snu.ac.kr}
}\vfil
\alignauthor{%
    \textbf{Sion Baek}\\
    \affaddr{LINE Plus}\\
    \affaddr{Seongnam, Republic of Korea} \\
    \email{sionbaek@linecorp.com}
}
\alignauthor{%
    \textbf{Bongwon Suh}\\
    \affaddr{Seoul National University} \\
    \affaddr{Seoul, Republic of Korea} \\
    \email{bongwon@snu.ac.kr}
}\vfil
\alignauthor{%
    \textbf{Joonhwan Lee}\\
    \affaddr{Seoul National University} \\
    \affaddr{Seoul, Republic of Korea} \\
    \email{joonhwan@snu.ac.kr}
}
}

\definecolor{linkColor}{RGB}{6,125,233}
\hypersetup{%
  pdftitle={\plaintitle},
  pdfauthor={\emptyauthor},
  pdfkeywords={\plainkeywords},
  bookmarksnumbered,
  pdfstartview={FitH},
  colorlinks,
  citecolor=black,
  filecolor=black,
  linkcolor=black,
  urlcolor=linkColor,
  breaklinks=true,
}

\begin{document}

%% For the camera ready, use the commands provided by the ACM in the Permission Release Form.
\CopyrightYear{2020}
\setcopyright{rightsretained}
\conferenceinfo{CHI '20 Workshop}{\emph{CHI 2020 Workshop on Conversational Agents for Health and Wellbeing}, April 26, 2020, Honolulu, HI, USA}
%% Then override the default copyright message with the \acmcopyright command.
\isbn{978-1-4503-6819-3/20/04}
\doi{}
\copyrightinfo{\acmcopyright}

\maketitle
% Uncomment to disable hyphenation (not recommended)
% https://twitter.com/anjirokhan/status/546046683331973120
\RaggedRight{} 

% Do not change the page size or page settings.
\begin{abstract}

Conversational agents have been showing lots of opportunities in healthcare by taking over a lot of tasks that used to be done by a human. One of the major functions of conversational healthcare agent is intervening users’ daily behaviors. In this case, forming an intimate and trustful relationship with users is one of the major issues. Factors affecting human-agent relationship should be deeply explored to improve long-term acceptance of healthcare agent. Even though a bunch of ideas and researches have been suggested to increase the acceptance of conversational agents in healthcare, challenges still remain. From the preliminary work we conducted, we suggest an idea of applying the personas of users’ family members and the doctor who are in the relationship with users in the real world as a solution for forming the rigid relationship between humans and the chatbot. 
\end{abstract}

\keywords{\plainkeywords}

% ACM Classfication

\begin{CCSXML}
<ccs2012>
<concept>
<concept_id>10003120.10003123</concept_id>
<concept_desc>Human-centered computing~Interaction design</concept_desc>
<concept_significance>500</concept_significance>
</concept>
</ccs2012>
\end{CCSXML}

\ccsdesc[500]{Human-centered computing~Interaction design}

% Print the classficiation codes
\printccsdesc

\section{Introduction}

Advances in voice recognition, natural language processing, and artificial intelligence have led to the increasing availability and use of conversational agents in the healthcare domain. The use of conversational agents is increasing not only in the medical organization but also in the field of daily health management. For example, intervening users’ behaviors that negatively affect their health status or provoking healthy behaviors that lead to positive health outcomes is becoming one of the major functions of conversational agents in healthcare. Chatbots have recently been highlighted as one of the most promising forms of healthcare system. ~\cite{8720289, 866bed09c5924f2ea388abf423447417}. This is because interventions for health management often rely on sense-making and learning, and conversation is an effective medium for serving this mechanism~\cite{cd3488862fa342378b3ba3c308572f9a}. For example, Huang et al~\cite{8607399} have used a chatbot for continuous management of user’s weight.

\begin{marginfigure}[-1pc]
  \begin{minipage}{\marginparwidth}
    \centering
    \includegraphics[width=1.0\marginparwidth]{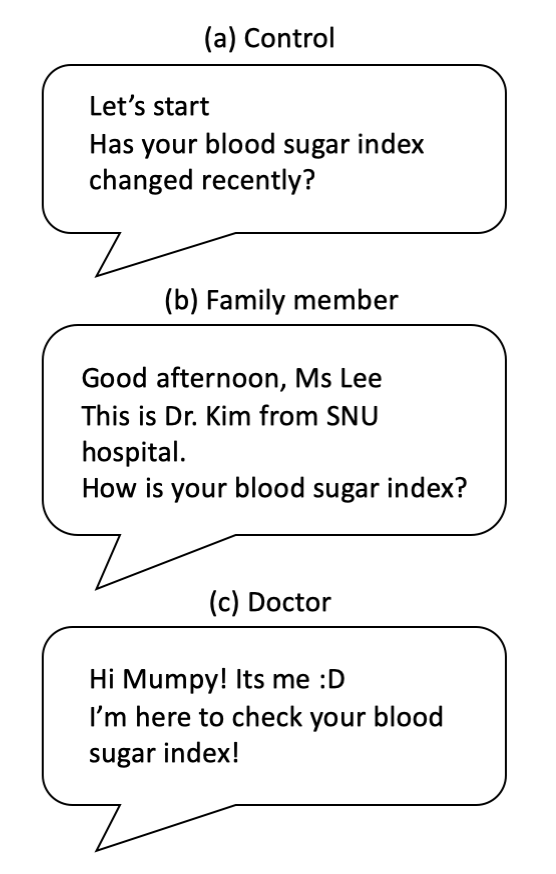}
    \caption{Sentences above are examples of conversational styles of three types of the chatbot. (a)Control chatbot uses a neutral tone of the conversation and it uses standard language style. (b)Chatbot with the persona of the user's family member was designed based on the previously made conversations between the user and the user's family member. It consists of at least 100 times of turn-taking. The conversational style of the (c)chatbot with the doctor's persona is implemented by the doctor himself. He transformed the series of sentences uses in the control chatbot into his own conversational style. }~\label{fig:boxplot}
  \end{minipage}
\end{marginfigure}

Managing daily behaviors that could potentially affect one’s health status that usually aim for long-term behavioral change. In this manner, forming an intimate, trustful relationship between the conversational agent and the users is essential and the relationship between them should be well designed based on the previous HCI works.
 
A lot of efforts were made to narrow down the relational gap between the healthcare agent and the user. When it comes to the conversational agent, the users’ perceived intimacy of the agent is influenced by the conversational style of the agent including tone of voice, linguistic markers, and frequent words, etc \cite{10.1371/journal.pone.0072589, Walters2008}. Integrating these factors, conversational agents’ particular persona perceived by the users affects the success of forming a rigid relationship in the human-agent interaction.
 
Applying persona to the agent has possibilities of enhancing a user's health management behavior by forming a positive relationship between them. Some relational agents have shown positive outcomes in healthcare by implementing a certain persona to the agent. Bickmore et al~\cite{10.1016/j.intcom.2009.12.001} found that agent with empathetic persona is effective for managing mental health while subtle persona agent instructing exercise increased the solidarity of the behavior change ~\cite{Bickmore03subtleexpressivity}. However, designing a persona that satisfies and every user is challenging since most of the systems with persona are not customized to individual user~\cite{Vandenberghe:2017:BPO:3027063.3052767}. This ends in a lack of intimacy between the system and the user in some cases. Moreover, generally designed personas even lead to a false sense of understanding. To prevent negative outcomes, applying the customized persona could be an ideal solution but few studies have dealt with this issue~\cite{Vandenberghe:2017:BPO:3027063.3052767}.

To design a customized persona for an individual's healthcare, we have applied the persona of users' support providers to the healthcare agent. We have developed and compared three chatbots. Two of them were chatbots with customized persona (users' family member and the doctor) and one was without persona (control). 

\section{Applying Persona of User's Support Providers to the Healthcare Chatbot}
In this position paper, we propose an idea of applying the persona of the user's support providers from the real world for designing customized persona of a healthcare agent. We first share results from our preliminary study. Then, we present the consequences of applying the persona of users' support providers who are in the relationship in the real world and discuss design considerations based on our results.

Among various types of support providers, we present the two example cases, the users' family member and the doctor that was applied to the chatbot. For persona development, differentiated appearance including profile image and profile name and conversational style were implemented. These features have been known to play a major role in persona perception~\cite{10.1371/journal.pone.0072589, Walters2008}.
To customize the appearance for both chatbots, profile pictures with a recognizable face was used as the agent’s profile (Figure 2) . For the conversational style, we collected conversation data from participant's family member and the doctor (Figure 1). All chatbots were built on KakaoTalk, the most popular messenger app in South Korea. 

\begin{marginfigure}[-3pc]
  \begin{minipage}{\marginparwidth}
    \includegraphics[width=1.0\marginparwidth]{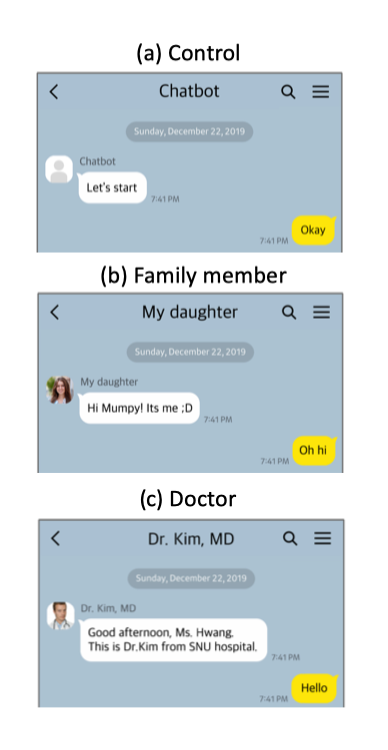}
    \caption{For persona perception, the appearance of a particular persona was implemented through the profile picture. For control chatbot, no profile picture was used to minimize the chance of persona perception by users. The assigned names of three chatbots were cautiously selected since it also largely impacts the possibilities of increasing persona perception. For control chatbot, we just named it chatbot. For the chatbot with the persona of the user's family member, it was named as the name of a particular person (e.g. James, Sarah, etc) or the relationship between them (e.g. my daughter, my mom, etc). }~\label{fig:boxplot}
  \end{minipage}
\end{marginfigure}

\section{The persona of users' family members and the doctor enhanced the relationship between the healthcare chatbot and the user.}

To evaluate three chatbots, we recruited eleven participants (Female: 6, Male: 5) aging from 19 to 60 years old (\textit{M} = 46 \textit{SD} = 13.82) with metabolic syndrome, since these people were highly in need of continuous behavior interventions~\cite{10.1210/er.2008-0024}. We conducted a within-subject study to measure the effects of personas on the human-chatbot relationship. Participants were instructed to interact with the chatbots with the given tasks related to daily behaviors that could potentially affect the user's health status. Tasks were related to diet, exercise, medication, alcohol consumption and smoking. They received behavioral intervention messages while interacting with chatbots. After completing all tasks, participants evaluated all chatbots with survey and semi-structured post-hoc interview. Questions they were asked included perceived intimacy and perceived trustfulness which is the key indicator of the rigid relationship \cite{reis1988intimacy}. Moreover, we added questionnaires evaluating system acceptance to answer if increased relationship positively affects system acceptance.  As a result, chatbots with the personas of users' family members and the doctor showed significantly higher perceived intimacy and the perceived trust than the control chatbot. Also, chatbots with the persona of their support providers were significantly more acceptable than the control chatbot with no application of persona. 

\textbf{Perceived intimacy of the healthcare chatbot with the family persona} Chatbot with the application of family member's persona showed significantly high perceived intimacy than the others (\textit{F}(2,30 = 19.421, \textit{p} <0.001)). From the interviews, we found most of the participants indicated the intimate relationship with the family persona affected the acceptance of the system. P6 said that "Chatbot mimicking My daughter is less annoying than any others because we are in a close relationship and I love her", P5 laughed and said that "I feel like I am really talking with my daughter." and P8 emphasized that "I know it is not my son, but I feel like I am talking with him which makes system friendly, and moreover, using this system will make me feel emotionally closer to him". In line with the interview results, perceived intimacy of healthcare chatbot with family persona was significantly related to system acceptance (\textit{Adj \(R^2\)} = 0.54,\textit{p} <0.001).

\textbf{Perceived trustfulness of the healthcare chatbot with the doctor's persona}
Chatbot with the persona of the user's personal doctor showed significantly high perceived trustfulness(\textit{F}(2,30 = 14.187, \textit{p} <0.001)) than any other healthcare chatbots. We found additional evidence from the interviews that perceived trustfulness of the chatbot with the doctor's persona was the key factor of healthcare chatbot's acceptance. For example, P10 said that "Even though the fundamental contents of the three systems are the same, more faith goes to an expert-like chatbot with a doctor's persona". Also, P9 mentioned that "Expert-like chatbot make the information seem more scientific and trustful which make the chatbot more acceptable".  In line with this interview results, perceived trustfulness of the healthcare chatbot with the doctor's persona was significantly related to system acceptance (\textit{Adj \(R^2\)} = 0.24,\textit{p} <0.01).

\section{Design considerations} 
Based on our preliminary results, we suggest design considerations for applying the user's support providers' persona to the healthcare chatbot. 
\newline
\textbf{Relational factors in the real world}
The relationship between the user and the healthcare chatbot with the persona of the user's support providers could largely be affected by the actual relationship between the user and the user's support provider in the real world. In our results, trustfulness had more influence on the human-agent relationship in the interaction between the user and the chatbot with the persona of the doctor than the persona of the family member. On the other hand, intimacy had more impact on the human-agent relationship in the interaction between the user and the healthcare chatbot with the persona of the family member than the doctor. From this, we could infer that defining relational factors derived from a relationship between the user and the actual support provider, whose persona is implemented to the system, has to be preceded before designing a healthcare chatbot with the persona of user's acquaintances. 

\textbf{Linguistic factors to consider when applying the persona of user's support providers to the healthcare chatbot}
The selection of a conversational style may contribute to designing a more acceptable healthcare chatbot. As indicated in our result, variation in the conversational style made participants perceive the agents more like their actual support providers. Since the ultimate goal of our agents with persona was to mimic positive aspects of the relationship between the user and the support providers, we carefully observed the effect of this facet as well. 

Specific linguistic factors played an important role in participants’ persona perception, making the agent more like the real family member and the doctor. General patterns of such differences are endearment (e.g. \textit{mumpy, sweetie, angel}), hedging (e.g. \textit{em, oh, ah}), frequent typos, word choice, and emojis. For example, P7 pointed out that \textit{“Using exclamation mark at the end of the sentence is what my sister is always doing! I feel like I am really having a conversation with her.”} P4 emphasized that hedging made the chatbot more like her son by saying that \textit{“My son always uses a word like `OMG' before he starts to say something. When I saw this word in the chatbot, I thought it really looked similar to a conversation with my son.”} However, a repetition of the same message content ends in negative consequences by making agents feel like a machine. P6, P9, and P10  mentioned that the repetition of the same contents will reduce the feeling of human likeness.

\section{Conclusion}
The novelty of our study lies in applying the traits of user's support providers to the healthcare chatbot for continuous and effective healthcare. While the potential of applying an actual person’s persona is yet to be explored in this field, progress in artificial intelligence and robotics is enabling computers to be more human-like. Computers could look and act like a specific individual in the near future. To take this opportunity, we have explored the opportunities of applying the personality traits of an actual person on the healthcare system.

\balance{} 

\bibliographystyle{SIGCHI-Reference-Format}
\bibliography{sample}

\end{document}